\documentclass[preprint2]{aastex}

\shorttitle{MAXIMUM BRIGHTNESS TEMPERATURE LIMIT}
\shortauthors{ASHOK K. SINGAL}

\begin{document}

\title{MAXIMUM BRIGHTNESS TEMPERATURE OF AN INCOHERENT SYNCHROTRON SOURCE : 
INVERSE COMPTON LIMIT - A MISNOMER}

\author{ASHOK K. SINGAL}
\affil{ASTRONOMY \& ASTROPHYSICS DIVISION, PHYSICAL RESEARCH LABORATORY, \\
NAVRANGPURA, AHMEDABAD - 380 009, INDIA; asingal@prl.res.in}

\begin{abstract}
We show that an upper limit of $\sim 10^{12} $~K on the peak brightness temperature 
for an incoherent synchrotron radio source,
commonly referred to in the literature as an inverse Compton limit,
may not really be due to inverse Compton effects. We show
that a somewhat tighter limit $T_{\rm eq} \sim 10^{11} $ is actually obtained 
for the condition of equipartition of energy between radiating particles and magnetic 
fields which happens to be a configuration of minimum energy 
for a self-absorbed synchrotron radio source.  An order of magnitude change in
brightness temperature from $T_{\rm eq}$ in either direction would require departures
from equipartition of about eight orders of magnitude, implying a change in
total energy of the system up to $\sim 10^{4} $ times the equipartition value. 
Constraints of such extreme energy variations imply that brightness 
temperatures may not depart much from $T_{\rm eq}$.
This is supported by the fact that at the spectral turnover, brightness temperatures 
much lower than $\sim 10^{11} $~K are also not seen in VLBI observations. Higher brightness 
temperatures in particular, would require in the source not only many orders of magnitude 
higher additional energy for the relativistic particles but also many order of magnitude 
weaker magnetic fields. Diamagnetic effects do not allow such extreme conditions, keeping 
the brightness temperatures close to the equipartition value, which is well below the 
limit where inverse Compton effects become important.

\end{abstract}

\keywords{galaxies: active --- quasars: general --- radiation mechanisms: 
non-thermal --- radio continuum: general}

\section{INTRODUCTION}
Kellermann \& Pauliny-Toth (1969) first suggested that the observed upper 
limit on the maximum radio brightness temperatures of compact self-absorbed radio 
sources is an inverse Compton limit. They argued that at brightness temperature 
$T_{\rm m} \stackrel{>}{_{\sim}}10^{12} $~K energy losses of radiating electrons due 
to inverse Compton effects become so large that these result in a rapid cooling
of the system, thereby bringing the synchrotron brightness temperature quickly below 
this limit. Singal (1986) on the other hand derived a somewhat tighter upper limit 
$T_{\rm m}\stackrel{<}{_{\sim}} 10^{11.5} $~K, without taking recourse to any inverse 
Compton effects. He used the argument that due to the 
diamagnetic effects the energy in the magnetic fields cannot be less than a certain 
fraction of that in the relativistic particles and then an upper limit on brightness 
temperature close to the equipartition value follows naturally. 
However Bodo, Ghisellini \& Trussoni (1992) pointed out that this limit on the 
magnetic field energy changes when the drift currents due to magnetic field gradients 
at the boundary are considered. Equipartition brightness temperature values have also 
been used to show that much higher Doppler factors are needed to successfully explain the 
variability events (Singal \& Gopal-Krishna 1985; Readhead 1994). 

Here we first examine the dependence of the brightness temperature on the magnetic 
field and relativistic particle energies and calculate the equipartition  
value $T_{\rm eq}$. We then show that how in a self-absorbed synchrotron case
$T_{\rm eq}$ corresponds to a minimum energy configuration for the source.
We further show why brightness temperatures of a source cannot rise much above this limit 
because of the diamagnetic effects, which keep the brightness temperatures close to the 
equipartition value, well below the limit where inverse Compton arguments become important.

Unless otherwise specified we use cgs system of units throughout.
\section{EQUIPARTITION BRIGHTNESS TEMPERATURE}
We want to examine an upper limit on the intrinsic brightness temperature 
achievable in a synchrotron source. Hence we will not consider here any effects
of the cosmological redshift or of the relativistic beaming due to a
bulk motion of the radio source, assuming that all quantities have been
transformed to the rest frame of the source.

In radio sources, specific intensity $I_{\nu}$, defined as the flux density
per unit solid angle at frequency $\nu$, usually follows a power law in the 
optically thin part of the spectrum, i.e., $I_{\nu} \propto \nu^{-\alpha} $,
arising from a power law energy distribution of radiating electrons 
$N(E)=N_{0} E^{-\gamma}$, with $\gamma = 2\alpha+1$.
The source may become self-absorbed below a turnover frequency $\nu_{\rm m}$
(see Pacholczyk 1970), where the peak intensity $I_{\rm m}$ is related 
to the magnetic field $B$ as,
\begin{equation}
B = 10^{-62}\, b(\alpha)\: \nu_{\rm m}^{5}\: I_{\rm m}^{-2}.
\end{equation}
Values of various parameters of spectral index $\alpha$, calculated from the 
tabulated functions in Pacholczyk (1977) are given in Table 1,
where we have made the plausible assumption that the direction of the
magnetic field vector, with respect to the line of sight, changes
randomly over regions small compared to a unit optical depth. 
Using Rayleigh-Jeans law,
\begin{equation}
I_{\rm m} = \frac{2\:k\: T_{\rm m}\, \nu_{\rm m}^{2}}{c^{2}}
= 3.07 \times 10^{-37}\: T_{\rm m}\, \nu_{\rm m}^{2},
\end{equation}
magnetic field $B$ can be expressed in terms of the peak brightness temperature $T_{\rm m}$,
\begin{equation}
B =  1.05 \times 10^{11}\, b(\alpha)\: \nu_{\rm m}\: T_{\rm m}^{-2}.
\end{equation}
The magnetic field energy density $W_{\rm b}= B^{2}/8 \pi$ can then be written as,
\begin{equation}
W_{\rm b}= 4.5 \times 10^{20} \, b^{2}(\alpha)\,\nu_{\rm m}^{2}\,T_{\rm m}^{-4}. 
\end{equation}
Energy density of the relativistic electrons in a synchrotron radio
source component is given by (Ginzburg \& Syrovatskii 1965; Ginzburg 1979)
\begin{eqnarray}
W_{\rm k}= \frac {1.48 \times 10^{12}} {s \:a(\alpha)} 
\frac {I_{\nu}\: \nu^{\alpha}\: B^{-1.5}} {(\alpha-0.5)}
\;\;\times  \;\;\;\;\;\;\;\;\;\;\;\;\; 
\nonumber \\
\left[\left(\frac {y_{1}(\alpha)}{\nu_{1}}\right) ^{\alpha-0.5} 
-\left(\frac {y_{2}(\alpha)}{\nu_{2}}\right) 
^{\alpha-0.5}   \right] , 
\end{eqnarray}
where $s$ is the characteristic depth of the component along the line of sight.
Using equations (2) and (3) we get,
\begin{eqnarray}
W_{\rm k}= \frac {1.3 \times 10^{-41}\: f(\alpha)} {s\: a(\alpha)\:(\alpha-0.5)
\,b(\alpha)^{1.5}} \;\nu_{\rm m}^{\alpha \,+\, 0.5}\,T_{\rm m}^{4}
\;\;\times  
\nonumber \\
\left[\left(\frac {y_{1}(\alpha)}{\nu_{1}}\right) ^{\alpha-0.5} 
-\left(\frac {y_{2}(\alpha)}{\nu_{2}}\right)^{\alpha-0.5} \right]. 
\end{eqnarray}
Writing $\eta = W_{\rm k}/W_{\rm b}$ we get,
\begin{equation}
\eta = \left(\frac{s}{\rm pc}\right)^{-1}\,
\, \left(\frac{\nu_{\rm m}}{\rm GHz}\right)^{\alpha-1.5}\,
\left(\frac{T_{\rm m}}{t(\alpha)\,10^{11}} \right)^{8}. 
\end{equation}
For typical values of $t(\alpha) \sim 0.6$ (Table 1), the denominator in the 
last term amounts to $\sim 10^{10.8}$.
Any variations in $\nu_{\rm m}$ that we may consider would at most be about an
order of magnitude around say, 1 GHz, (after all it is the brightness temperature 
limits seen in the radio-band that we are trying to explain), 
which will hardly affect $T_{\rm m}$ (Equation 7). For example, for typical $\alpha$
values of 0.75 to 1.0, a factor of 10 change in $\nu_{\rm m}$ will change $T_{\rm m}$ 
by a factor $\stackrel{<}{_{\sim}}(10)^{0.1}$ . Further, 
with the reasonable assumption that a self-absorbed radio source size may not be much 
larger than $\sim$ a pc, we see that at equipartition ($\eta = 1$) the brightness 
temperature value is $T_{\rm eq}\stackrel{<}{\sim} 10^{11}$.

\section{A CORRECTION TO THE DERIVED $T_{\rm m}$ VALUES}
Actually the $T_{\rm m}$ values that have been considered 
hitherto (here as well as in the past literature) are for the turnover point in the 
synchrotron spectrum where the flux density peaks. However, a maxima of flux density 
is not necessarily a maxima for the brightness temperature as well since the definition 
of the brightness temperature also involves $\nu^{2}$ (Equation 2). 
In fact a zero slope for the flux density with respect to $\nu$ would imply for the 
brightness temperature a slope of $-2$. Therefore the peak of the brightness temperature 
will be at a point where flux density $\propto \nu^{2}$, so that 
$T_{b} \propto F \, \nu^{-2}$ has a zero slope with respect to $\nu$.  The peak of
$F_{\nu} \, \nu^{-2}$, can be determined in the following manner (Singal 2009).  The
specific intensity in a synchrotron self-absorbed source is given by,
\begin{eqnarray}
I_{\nu} = \frac {c_{5}(\alpha)}{c_{6}(\alpha)} \left(\frac {\nu}{2c_{1}}\right)^{2.5} 
\:B_{\perp}^{-0.5}
\;\;\times  \;\;\;\;\;\;\;\;\;\;\;\;\; 
\nonumber \\
\left[1-\exp\left\{-\left( \frac{\nu}{\nu_{1}}\right)^{-(\alpha+2.5)}\right\}\right] , 
\end{eqnarray}
where $c_{1}, c_{5}(\alpha), c_{6}(\alpha)$ are tabulated in Pacholczyk (1970). 
The optical depth varies with frequency as $\tau=(\nu/\nu_{1})^{-(\alpha+2.5)}$,  
$\nu_{1}$ being the frequency  at which $\tau$ is unity. The equivalent brightness 
temperature (in Rayleigh-Jeans limit) is then given by
\begin{eqnarray}
T_{\nu} = \frac {c^{2}\, c_{5}(\alpha)}{8\, k\, c_{1}^{2}\, c_{6}(\alpha)} 
\left(\frac {\nu}{2c_{1}}\right)^{0.5} \:B_{\perp}^{-0.5}
\;\;\times  \;\;\;\;\;\;\;\;\;\;\;\;\; 
\nonumber \\
\left[1-\exp\left\{-\left( \frac{\nu}{\nu_{1}}\right)^{-(\alpha+2.5)}\right\}\right] .
\end{eqnarray}
We can maximize $T_{\nu}$ by differentiating it with $\nu$ and equating
the result to zero. This way we get an equation for the optical depth
$\tau_{\rm o}$, corresponding to the {\em peak brightness temperature}
$T_{\rm o}$, which is different from the one that is available in the
literature for the optical depth $\tau_{\rm m}$ at the {\em peak of 
the spectrum}.  The equation that we get for $\tau_{\rm o}$ is
\begin{equation} 
\exp\,(\tau_{\rm o})=1+(2\alpha+5)\,\tau_{\rm o}.
\end{equation}
Solutions of this transcendental equation for various $\alpha$
values are given in Table 1. It is interesting to note that while the peak of the spectrum 
for the typical $\alpha$ values usually lies in the optically thin part of the spectrum 
($\tau_{\rm m}\stackrel{<}{_{\sim}}1$; Table 1), peak of the brightness
temperature lies deep within the optically thick region ($\tau_{\rm o}\sim 3$). 
Both the frequency and the intensity have to be calculated for $\tau_{\rm o}$ to get 
the maximum brightness temperature values. The correction factors are then given by,
\begin{equation}
\frac{\nu_{\rm o}}{\nu_{\rm m}}=\left(\frac{\tau_{\rm m}}{\tau_{\rm o}}\right)^{1/(\alpha+2.5)}
\end{equation}
\begin{equation}
\frac{T_{\rm o}}{T_{\rm m}}=\left(\frac{\nu_{\rm o}}{\nu_{\rm m}}\right)^{0.5}\,
\left[\frac {1-\exp\,(-\tau_{\rm o})}{1-\exp\,(-\tau_{\rm m})}\right]
\end{equation}
In Table 1 we have listed $\nu_{\rm o}/\nu_{\rm m}$ for various $\alpha$ values.  
The corresponding correction factors ($T_{\rm o}/T_{\rm m}$) to the peak brightness 
temperature values, which need to be applied in both inverse Compton 
and equipartition cases before making a comparison with the observed values, 
are given in Table 1. Also listed are the accordingly corrected $T_{\rm eq}$ values. 
From observational data, the deduced values (Readhead 1994; Homan et al. 2006) of the 
intrinsic brightness temperature are $T_{\rm b}$ $\stackrel{<}{_{\sim}} 10^{11.3} $~K,
quite consistent with the $T_{\rm eq}$ values from Table 1.

\section{EQUIPARTITION AND MINIMUM ENERGY DENSITY}
We notice that for any increase in the brightness temperature at any given turnover frequency, 
while the energy density of radiating particles has to go up by a factor $\propto T_{\rm m}^{4}$ 
(Equation 6), that in the magnetic fields will have to go down by a similar factor (Equation 4). 
We can then derive a minimum energy density of the system. The total energy density 
$W_{\rm k}+W_{\rm b}=c_{1}T_{\rm m}^{4}+c_{2}T_{\rm m}^{-4}$ can be minimized 
with respect to $T_{\rm m}$ to get $W_{\rm k}=W_{\rm b}$ as the condition for the 
minimum total energy density. It should be noted that the relation between 
$W_{\rm k}$ and $W_{\rm b}$ here is somewhat different from the case of extended sources, 
where we get minimum energy density for an approximate equipartition condition 
$W_{\rm k}=\frac{4}{3} W_{\rm b}$ (see Pacholczyk 1970). The reason being that in
extended sources the intensity was treated as an independent quantity, to be determined 
from observations, while in a self-absorbed case the maximum intensity and therefore 
$T_{\rm m}$ is tied to the magnetic field value through equations (1~and~3) 
yielding an exact equipartition condition. The equipartition brightness temperature 
$T_{\rm eq}\sim 10^{11}$ thus corresponds to a minimum energy 
configuration of the system. It follows from Equation (7) that an order of magnitude
higher $T_{\rm m}$ values would require $\eta$ to increase by about a
factor ${\sim} 10^{8}$, i.e., departure from equipartition will go
up by about eight orders of magnitude.  Actually for a given $\nu_{\rm m}$, 
the magnetic field energy density will go down by a factor
$\sim 10^{4}$ (Equation 4), while that in the relativistic particles will
go up by a similar factor (Equation 6).  This implies that the total 
energy budget of the source will also need be higher by about $\sim 10^{4}$ than 
from the minimum energy value. 

Constraints of such extreme energy variations 
with brightness temperature imply that the latter may not depart much from 
$T_{\rm eq}$. This is supported by the fact that brightness temperatures 
much lower than $\sim 10^{11} $~K are also not seen at the spectral peak
(see e.g., Kellermann \& Pauliny-Toth 1969, Fig. 4), since brightness temperatures 
much lower than $T_{\rm eq}$ as well require much larger total energies. 
(It is to be emphasized that the brightness temperatures being considered here are 
the peak  values near the turnover in the synchrotron self-absorbed sources, and not 
the lower values which in any case occur in the optically thin regions). 

\section{DIAMAGNETIC EFFECTS}

How high could the brightness temperature rise above the equipartition value in a 
synchrotron source? Following Homan et al. (2006) we can envisage a scenario in which the 
particle energy density is increased by injecting a large number of  additional relativistic 
particles into the system, e.g., by increasing $N_{0}$, as $W_{\rm k} \propto N_{0}$
for any given energy index $\gamma$. $T_{\rm m}$ remains below $T_{\rm eq}$ as long as 
$W_{\rm k}<W_{\rm b}$. To increase $T_{\rm m}$ further, the required change in $N_{0}$ 
can be more conveniently calculated from a proportionality expression that can be derived 
directly from synchrotron self-absorption (Pacholczyk 1970),
\begin{equation}
T_{\rm m} \propto \left(\frac{\nu_{\rm m}}{B}\right)^{0.5} 
\propto \left(\frac{N_{\rm 0}}{B}\right)^{1/(2 \alpha + 5)} .
\end{equation}
Now an order of magnitude change in $T_{\rm m}$ for a given magnetic field $B$ will require 
the particle energy density to increase $10^{6} - 10^{7}$ times. However any change 
in $W_{\rm k}$ also brings a change in $B$ due to diamagnetic effects. Gyrating charged 
particles create their own magnetic field, in a direction opposite to the original field, 
thus giving rise to diamagnetic effects (Ginzburg \& Syrovatskii 1969; Singal 1986). 
If $H$ is the original magnetic field  (that is, the magnetic field value in the 
absence of diamagnetic effects) then the resultant field is given by (Bodo et al. 1992),
\begin{equation}
\frac{W_{\rm k}}{3} + \frac{B^{2}}{8 \pi} = \frac{H^{2}}{8 \pi} .
\end{equation}
This equation actually states that the pressure due to particles and magnetic field in the 
inner regions is balanced by the surrounding field pressure (Schmidt 1979). At equipartition, 
energy density of radiating particles and magnetic fields is equal, 
$W_{\rm k}=W_{\rm b}=W_{\rm eq}$ (say), which implies $W_{\rm eq}=\frac{3}{4} H^{2}/8 \pi$. 
We can then rewrite Equation (14) as
\begin{equation}
\frac{W_{\rm k}}{3} + \frac{B^{2}}{8 \pi} = \frac{4\,W_{\rm eq}}{3} .
\end{equation}
Due to diamagnetic effects an increase in $W_{\rm k}$ not only lowers $W_{\rm b}$, 
but it may also shift the turnover frequency somewhat 
since from equations (4 and 6) $W_{\rm k} W_{\rm b} \propto \nu_{\rm m}^{\alpha \,+\, 2.5}$. 
Apart from equipartition, another solution of Equation (15) exists for $W_{\rm k}= 3 W_{\rm eq}$  
and $W_{\rm b}=W_{\rm eq}/3$ which does not change $\nu_{\rm m}$.  However the resultant increase in 
$T_{\rm eq}$ from Equation (7) is only a factor of $9^{1/8} \sim (10)^{0.1}$.
The maximum $W_{\rm k}$ that can be achieved is only $4 W_{\rm eq}$. Larger 
$W_{\rm k}$ may lead to a total screening of the field, instabilities or other circumstances 
like total disruption of the source as the external field pressure $H^{2}/8 \pi$ may not
be able to contain the inner pressure, or the source might expand adiabatically to find a 
situation which is not too far from equipartition (Ginzburg \& Syrovatskii 1969; 
Bodo et al. 1992). Except in such non-equilibrium conditions, where large amount of particle
energy might have been injected (e.g., near the base of the radio jets) and the system has not  
yet relaxed to equilibrium, the results derived here should hold good.

\section{DISCUSSION AND CONCLUSIONS}
Kellermann \& Pauliny-Toth (1969) explained the observed upper limit on the 
radio brightness temperatures of compact self-absorbed radio sources in terms of 
Inverse Compton losses. They derived the ratio of inverse Compton to synchrotron 
radiation losses as,
\begin{eqnarray}
\frac {P_{\rm c}}{P_{\rm s}}=\frac {W_{\rm p}}{W_{\rm b}} \sim \left(\frac{\nu_{\rm m}}
{\rm GHz}\right) \left(\frac{T_{\rm m}}{10^{11.5}}\right)^{5}\;\;\times  
\;\;\;\;\;\;\;\;\;\;\;\;\; 
\nonumber \\
 \left[1+\left(\frac{\nu_{\rm m}}
{\rm GHz}\right) \left(\frac{T_{\rm m}}{10^{11.5}}\right)^{5}\right], 
\end{eqnarray}
here $W_{\rm p}$ is the synchrotron photon energy density and the second term represents 
the effect of the second-order scattering. From this
we gather that the two rates are comparable at $T_{\rm m}\sim 10^{11.5}$~K. 
At higher brightness temperatures, say at 
$T_{\rm m} \stackrel{>}{_{\sim}}10^{12} $~K, energy losses of radiating 
electrons due to inverse
Compton effects would become extremely large, resulting in a rapid cooling
of the system and thereby bringing the synchrotron brightness temperature
quickly below this value. However, these
inverse Compton losses become important {\it only if} $T_{\rm m} \stackrel{>}{\sim}
10^{12} $~K. It should be noted that even though inverse
Compton scattering increases the photon energy density, yet it does
not increase the radio brightness as the scattered photons get boosted
to much higher frequencies (in range of infrared to X-rays). 
If anything, some photons get removed
from the radio band, but the change in radio brightness due to that itself
may not be very large.  What could be important is the large energy
losses by electrons which may cool the system rapidly. 
But can brightness temperatures ever rise to such high
values for inverse Compton losses to come into play at a significant level?

The equipartition conditions may keep the temperatures well below this limit,
as departures from minimum energy configuration may not grow very large. 
This is not to say that inverse Compton effects cannot occur; it is only that 
conditions in synchrotron radio sources may not arise for inverse Compton 
losses to become very effective. Here it may appear curious that two apparently different 
theoretical approaches lead to brightness temperature limits which are rather similar;
inverse Compton value being only about three times higher than the equipartition one. 
The genesis of this similarity lies in the fact that for any given turnover frequency,  
$T_{\rm m}$ is exclusively determined by the magnetic field value (Equation 3), 
and that in both cases, a common factor $T_{\rm m}^{4}$ arises due to $W_{\rm b}$ 
in the denominator. 

But can magnetic fields get low enough for $T_{\rm m}$ to rise significantly where
inverse Compton effects become appreciable? In the case $W_{\rm k}$ approaches $4W_{\rm eq}$, 
from Equation (15) $W_{\rm b}$ could become very small, thereby making $\eta$ extremely large.  
However in this case the turnover frequency $\nu_{\rm m}$ will also shift to a very low value. 
An order of magnitude increase in $T_{\rm m}$ from Equation (13) will imply magnetic field $B$ 
falling by a factor $\sim 10^{-6}$ (i.e., $W_{\rm b}$ falling by $\sim 10^{-12}$) and 
$\nu_{\rm m}$ falling by $\sim 10^{-4}$. Even if such an unrealistic drop in magnetic field 
value to nano-Gauss and turnover frequency to kHz range were achievable, for one thing 
this will take $\nu_{\rm m}$ outside the radio-band of our interest where the brightness 
temperature limits have been seen and which we are trying to explain here. But even more 
important the consequential increase in brightness temperature will still be not contained 
by inverse Compton effects as even the inverse Compton limit will go up about an order of 
magnitude for such a large fall in the turnover frequency (Equation 16). Hence, it is 
imperative that even in such a scenario inverse Compton effects do not get to play any 
significant role in maintaining the maximum brightness temperature limit in an incoherent 
synchrotron radio source.

What about the effects of inhomogeneities in the source on the brightness temperature limits? 
As mentioned in Section 2, we have assumed that the direction of the magnetic field vector, 
with respect to the line of sight, changes randomly over the source. This makes $a$
in Equation~(6) change by a factor $\sim 0.67$ (Ginzburg 1979), while $b$ in Equation~(3) 
changes by a factor $\sim 1.15$ ($=(c_{29}\, c_{14})^{2}$ from Tables of Pacholczyk 1970, 1977). 
From Equations (4, 6 and 7) we see that $T_{\rm m} \propto [a\, b^{3.5}]^{1/8}$. 
Thus our derived $T_{\rm m}$ values in the case of a random magnetic field orientation are higher 
than those in the homogeneous case by a factor of $1.01 \sim (10)^{0.005}$, a negligible quantity. 
Another type of inhomogeneity one could consider is when the source may consist of a number of 
discrete components, each with its own cutoff frequency; such a scenario is suggested by the flat 
shape of the observed spectra as well as the VLBI observations (Kellermann \& Pauliny-Toth 1981).
However in such a case the observed brightness temperature value, a sort of average
over the various components, cannot exceed the highest value among the individual components,  
which will have an equipartition brightness temperature limit as derived above. 

We can thus conclude that under a variety of conditions the diamagnetic effects will limit the 
brightness temperatures close to the equipartition value, well below the limit where inverse 
Compton effects become important.
%------------------------------------------------------------------
%\clearpage
\begin{table}[t]
\begin{center}
\caption{Functions of the Spectral Index $\alpha$}
\begin{tabular}{lclccclcllccc}
\tableline\tableline
$\,\,\alpha$ &  $\gamma$ & $ \,\,a(\alpha)$ & $b(\alpha)$ & $f(\alpha)$  & $y_{1}(\alpha)$ 
& $y_{2}(\alpha)$&  $t(\alpha)$\tablenotemark{a} & $\;\tau_{\rm m}\;\;$ 
& $\;\tau_{\rm o}$  & $\nu_{\rm o}/\nu_{\rm m}$ & $T_{\rm o}/T_{\rm m}$ & log $T_{\rm eq}$\\
\noalign{\smallskip}\hline\noalign{\smallskip}
0.25 & 1.5 & 0.149  & 0.61 & 1.10 & 1.3 & 0.011 & 0.53 & 0.19 & 2.80 & 0.37 & 3.5 & 11.3\\
0.5  & 2.0 & 0.103  & 0.85 & 1.19 & 1.8 & 0.032 & 0.67 & 0.35 & 2.92 & 0.50 & 2.2 & 11.2\\
0.75 & 2.5 & 0.0831 & 0.84 & 1.27 & 2.2 & 0.10  & 0.63 & 0.50 & 3.03 & 0.58 & 1.8 & 11.1\\
1.0  & 3.0 & 0.0741 & 0.74 & 1.35 & 2.7 & 0.18  & 0.53 & 0.64 & 3.13 & 0.64 & 1.6 & 10.9\\
1.5  & 4.0 & 0.0726 & 0.52 & 1.50 & 3.4 & 0.38  & 0.34 & 0.88 & 3.32 & 0.72 & 1.4 & 10.7\\
\tableline
\end{tabular}
\tablenotetext{a}{for $\nu_{1}$ and $\nu_{2}$ taken as 0.01 and 100 GHz respectively.}

\end{center}
\end{table}
%------------------------------------------------------------------

%\clearpage


\begin{thebibliography}{99.}
\bibitem{}Bodo, G., Ghisellini, G., \& Trussoni, E. 1992, MNRAS, 255, 694 
\bibitem{}Ginzburg, V. L. 1979, Theoretical Physics and Astrophysics (Oxford: Pergamon)
\bibitem{}Ginzburg, V. L., \& Syrovatskii, S. I. 1965, ARA\&A, 3, 297  
\bibitem{}Ginzburg, V. L., \& Syrovatskii, S. I. 1969, ARA\&A, 7, 375  
\bibitem{}Homan, D. C. et al 2006, ApJ, 642, L115  
\bibitem{}Kellermann,  K. I., \& Pauliny-Toth, I. I. K. 1969, ApJ, 155, L71
\bibitem{}Kellermann,  K. I., \& Pauliny-Toth, I. I. K. 1981, ARA\&A, 19, 373
\bibitem{}Pacholczyk, A. G. 1970, Radio Astrophysics (San Francisco: Freeman)
\bibitem{}Pacholczyk, A. G. 1977, Radio Galaxies (Oxford: Pergamon)
\bibitem{}Readhead, A. C. S. 1994, ApJ, 426, 51
\bibitem{}Schmidt, G. 1979, Physics of High Temperature Plasma (New York: Academic) 
\bibitem{}Singal, A. K. 1986, A\&A, 155, 242
\bibitem{}Singal, A. K. 2009, in Ap\&SS Proc., Turbulence, Dynamos, Accretion Disks, 
Pulsars and Collective Plasma Processes, ed. S. S. Hasan, R. T. Gangadhara, \& V. Krishan  
 (Berlin: Springer), 273
\bibitem{}Singal, A. K., \& Gopal-krishna 1985, MNRAS, 215, 383
\end{thebibliography}
\end{document}